\documentclass{PoS}

\usepackage{graphicx}
\usepackage{graphics}
\usepackage{txfonts}
\usepackage{epsfig}
\usepackage{natbib}
\usepackage{times}
\usepackage{amssymb}

\newcommand\ergcms{erg\,cm$^{-2}$\,s$^{-1}$}

\newcommand\ergs{erg\,s$^{-1}$}
\newcommand\cmsq{cm$^{-2}$}
\newcommand\integ{{\it{INTEGRAL}}}
\newcommand\suz{{\it{Suzaku}}}

\newcommand\swift{{\it{Swift}}}

\newcommand\xmm{{\it{XMM-Newton}}}

\newcommand\nh{$N_\mathrm{H}$}

\title{\suz\ captures a possible eclipse in IGR~J16207$-$5129 and identifies a weak-flaring state in IGR~J17391$-$3021}

\ShortTitle{\suz\ observations of IGR~J16207$-$5129 and IGR~J17391$-$3021}

\author{\speaker{Arash Bodaghee}%\\%
        \thanks{AB and JAT acknowledge grants: \emph{Suzaku} NNX08AB88G, \emph{Chandra} G089055X, and \emph{INTEGRAL} NNX08AC91G.} \\
       Space Sciences Laboratory, University of California, Berkeley \\
       E-mail: \email{bodaghee@ssl.berkeley.edu}}

\author{John A. Tomsick\\
        Space Sciences Laboratory, University of California, Berkeley}

\author{J\'er\^ome Rodriguez\\
        CNRS/INSU/AIM/IRFU/DSM/SAp, CEA Centre de Saclay, Universit\'e Paris Diderot}

\author{Sylvain Chaty\\
        CNRS/INSU/AIM/IRFU/DSM/SAp, CEA Centre de Saclay, Universit\'e Paris Diderot}

\author{Katja Pottschmidt\\
	CRESST, NASA Goddard Space Flight Center}

\author{Roland Walter\\
	ISDC, Observatoire de l'Universit\'e de Gen\`eve}

\author{Patrizia Romano\\
	INAF, Istituto di Astrofisica Spaziale e Fisica Cosmica}

\abstract{We present the results from analyses of \suz\ observations of the supergiant X-ray binaries IGR J16207$-$5129 and IGR J17391$-$3021. For IGR J16207$-$5129, we provide the first broadband (0.5--60 keV) spectrum from which we confirm a large intrinsic column density (\nh\ $= 16\times10^{22}$ \cmsq), and constrain the cutoff energy for the first time ($E_{c} = 19$ keV). We observed a prolonged ($>$ 30 ks) attenuation of the X-ray flux which we tentatively attribute to an eclipse of the probable neutron star by its massive companion. For IGR J17391$-$3021, we witnessed a transition from quiescence to a low-activity phase punctuated by weak flares whose peak luminosities in the 0.5--10 keV band are only a factor of 5 times that of the pre-flare emission. The weak flaring is accompanied by an increase in the absorbing column which suggests the accretion of obscuring clumps of wind. Placing this observation in the context of the recent \swift\ monitoring campaign, we now recognize that these low-activity epochs constitute the most common emission phase for this system, and perhaps in other SFXTs as well.}

\FullConference{8th INTEGRAL Workshop ``The Restless Gamma-ray Universe'' \\
		 September 27-30 2010\\
		 Dublin Castle, Dublin, Ireland}

\begin{document}

\section{Introduction \& Observations}

Among the more intriguing results by \integ\ has been the discovery of new types of high-mass X-ray binaries (HMXBs) with supergiant companions. They can be broadly divided into systems that are persistently-emitting (SGXBs: supergiant X-ray binaries) and those that emit sporadically (SFXTs: supergiant fast X-ray transients). The distinction between the two groups is not as clearly defined as once believed. Therefore, the continued study of these systems is crucial for understanding the origin and evolution of these previously-rare HMXBs.

\begin{figure*}[!t] 
\vspace{-5mm}
\includegraphics[width=\textwidth,angle=0]{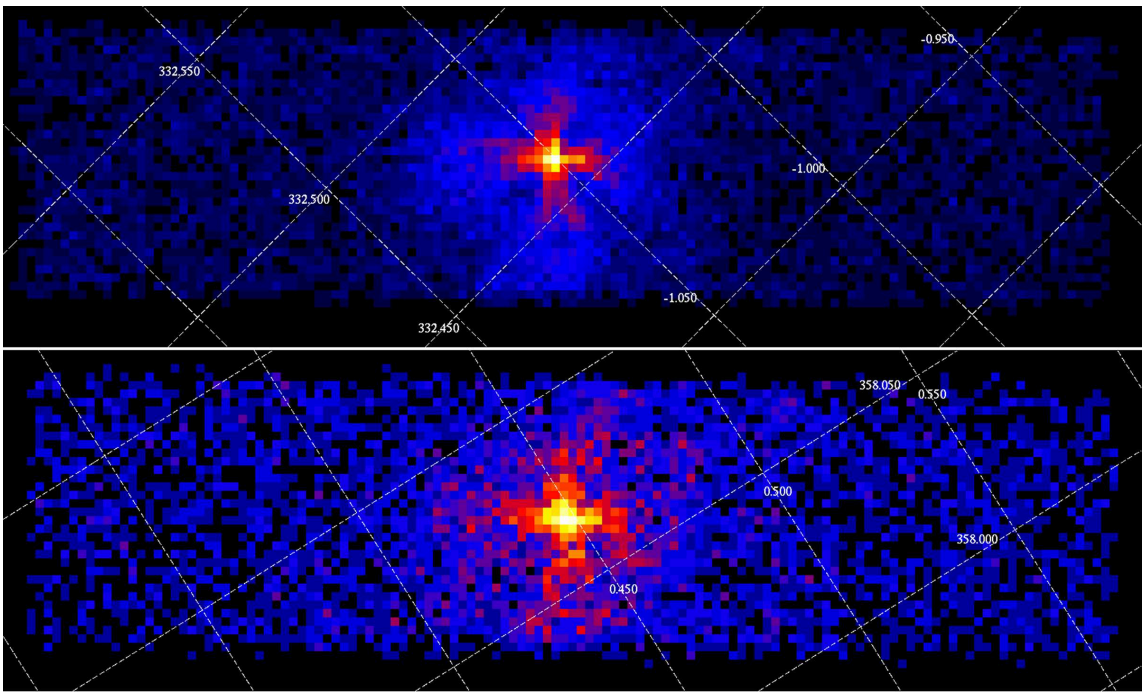}
\caption{Intensity images in Galactic coordinates of IGR~J16207$-$5129 (top) and IGR~J17391$-$3021 (bottom) in the 0.5--10 keV energy range as taken by the XIS-1 detector aboard \suz. }
\label{fig_img}
\end{figure*}

In 2008, the \suz\ space telescope observed two such objects: the SGXB named IGR~J16207$-$5129 and the SFXT dubbed IGR~J17391$-$3021 (= XTE~J1739$-$302). Fig. \ref{fig_img} presents images of the fields around the 2 sources. Note that XIS did not record data during the first observation of IGR~J16207$-$5129. In addition, IGR~J17391$-$3021 was very faint during its observation so HXD recorded few photon counts. Hence, HXD data were ignored in the analysis of IGR~J17391$-$3021. After removing bad events, and after extracting gaps in the observation, the effective exposure times that remained were 33 and 37 ks for IGR~J16207$-$5129 and IGR~J17391$-$3021, respectively.

%__________________________________________________________________XIS Image

%%__________________________________________________________________Journal
%%
%\begin{table*}[!t] 
%%\begin{scriptsize}
%\begin{tabular}{ l l l c }
%\hline
%\hline
%source			& observation ID      			& instruments			& dates				\\
%				&						& 					& (MJD)				\\	
%\hline
%IGR~J16207$-$5129	& 402065010 ($\equiv$ O1) 	& HXD 				& 54499.823--54500.784	 \\	
%				& 402065020 ($\equiv$ O2) 	& HXD $+$ XIS			& 54526.866--54527.750	 \\	
%\hline
%IGR~J17391$-$3021	& 402066010			 	& HXD $+$ XIS			& 54518.495--54519.396	 \\	
%\hline
%\end{tabular}
%\caption{Journal of \suz\ observations of the HMXBs in this study.}
%\label{tab_log}
%%\end{scriptsize}
%\end{table*}

\section{X-ray Variability}

%__________________________________________________________________LC
\begin{figure*}[!t] 
\includegraphics[width=7.7cm,angle=0]{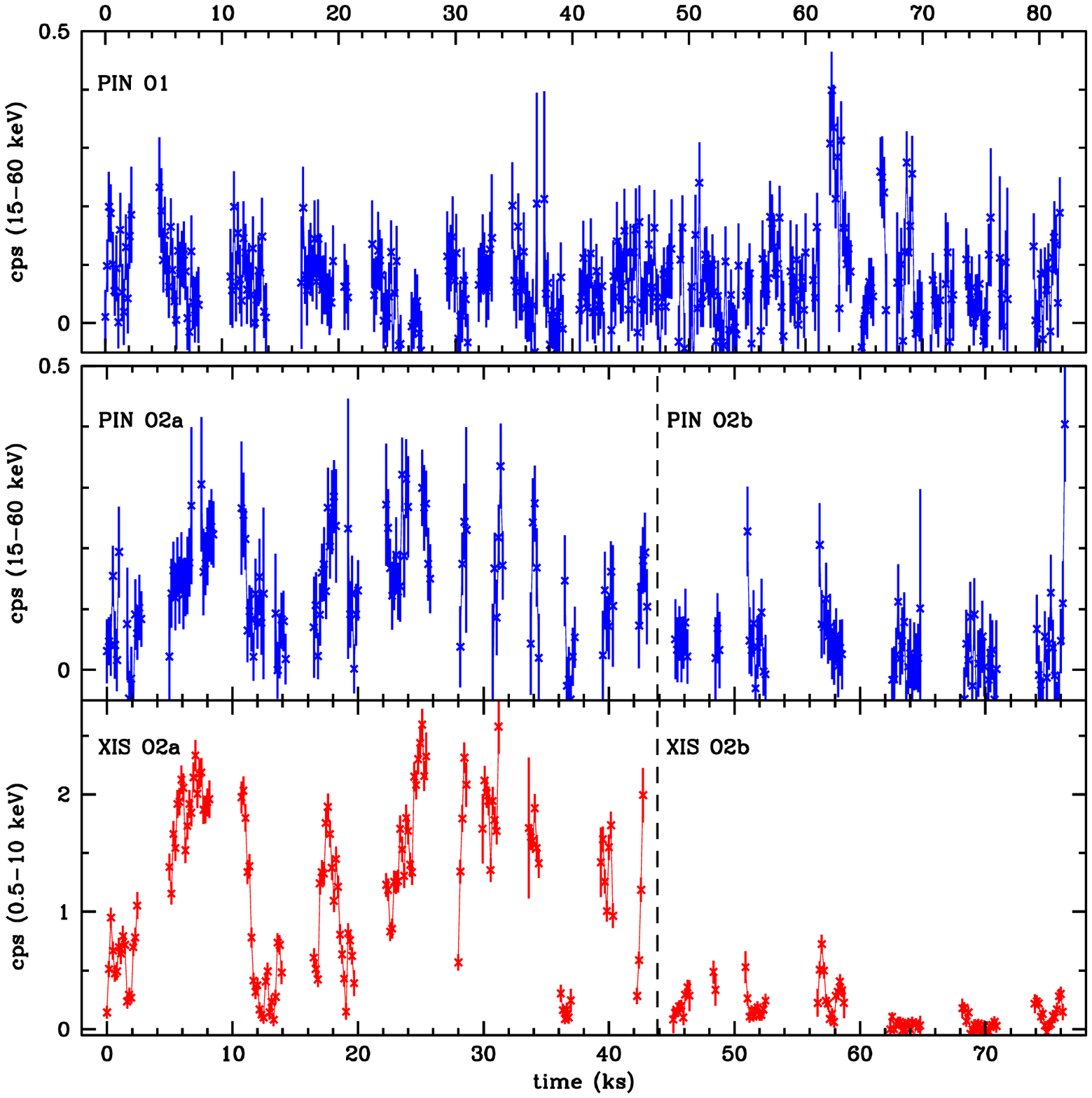}\includegraphics[width=7.7cm,angle=0]{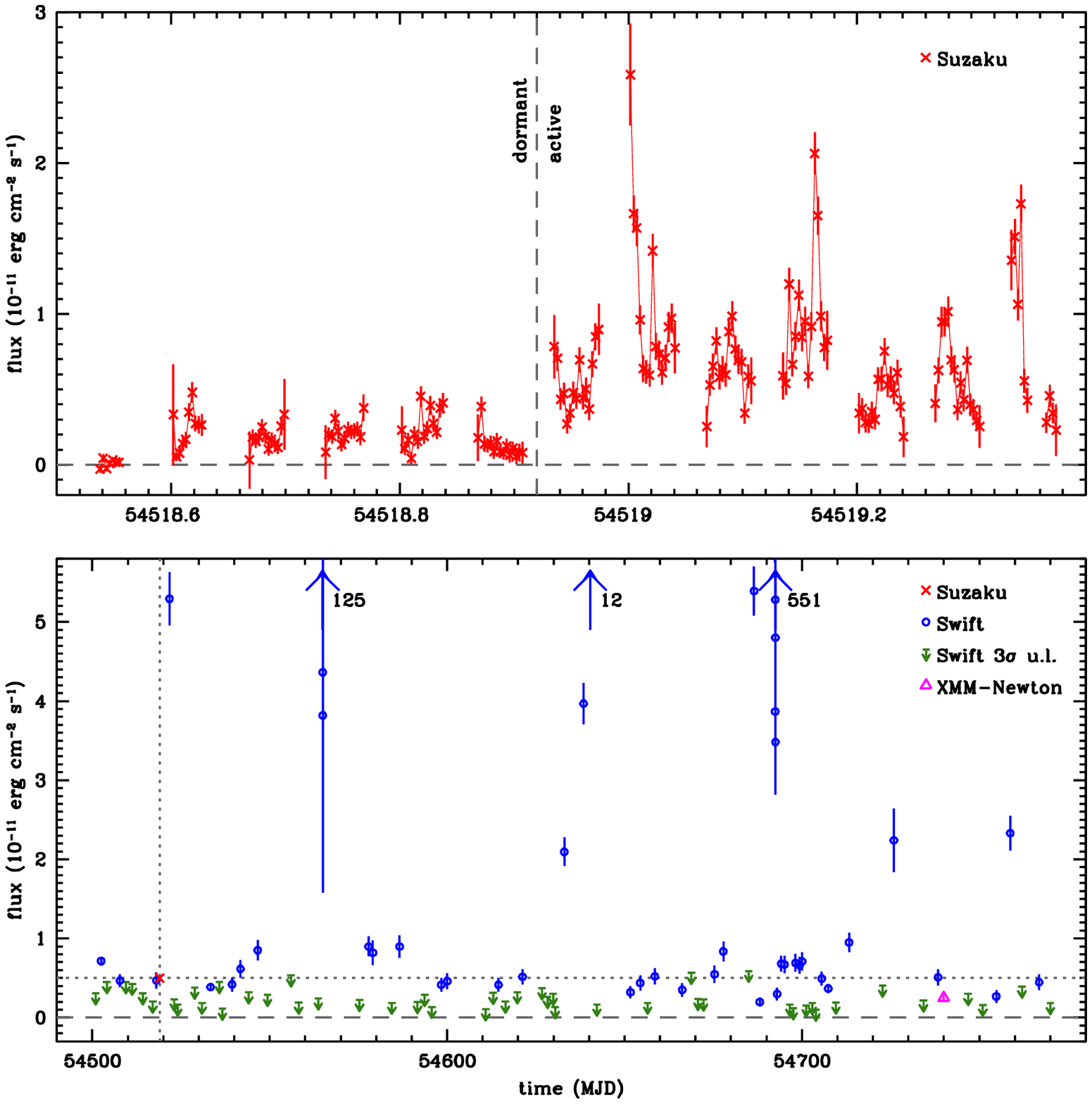}
\caption{\emph{Left}: Background-subtracted light curve of IGR~J16207$-$5129 from XIS (red: 0.5--10 keV) and HXD (blue: 15--60 keV). The upper panel displays the first observation (O1) while the lower panel shows the second observation (O2) which includes simultaneous data from XIS. Each bin collects 160 s worth of data. The dashed line represents MJD 54527.382. This corresponds to the onset of an unusually long period (more than 30 ks) of suppressed flux which could represent an eclipse of the compact object by its supergiant stellar companion. \emph{Right}: Light curve of IGR~J17391$-$3021 from XIS, \swift-XRT, and \xmm. The top panel presents the background-subtracted light curve from our \suz\ observation (red crosses, 240-s binning). In the bottom panel, the source light curve from the Swift monitoring campaign of \citet{Rom09} is shown (blue circles, $\sim$1 ks of exposure time per data point) along with their 3-$\sigma$ upper limits for non-detections (green downward arrows). The blue upwards arrows designate the maximum flux of large outbursts detected by \swift\ that are situated beyond the scale of the graph. The average flux from our 37-ks \suz\ observation is plotted as a single red cross at the intersection of the dotted lines. The magenta triangle at MJD 54740 corresponds to the average flux in a 31-ks observation with \xmm\ \citep{Boz10}. Fluxes are given in the 0.5--10 keV energy band as observed in units of $10^{-11}$ \ergcms.}
\label{fig_lc}
\end{figure*}

\subsection{IGR~J16207$-$5129}

The light curve from IGR~J16207$-$5129 (Fig. \ref{fig_lc}) illustrates that the emission varies within 1--2 orders of magnitude on ks timescales, which is typical of persistent SGXBs. The maximum luminosity is $5\times10^{35}$ \ergs\ ($d$/6 kpc)$^{2}$, while the minimum is $2\times10^{34}$ \ergs. 

Intriguingly, X-rays from the source, and the variability associated with that emission, are significantly diminished during the last $\sim$30 ks of the observation. Several mechanisms can be invoked to explain the suppression of X-ray emission for a sustained period. If the accretor were to enter an ``off'' state, this would last a few 100 s, or at least an order of magnitude shorter than the period of inactivity that we observed. Occulting clumps in the wind or hydrodynamic effects would lead to changes in \nh, which we do not see. An occultation of the neutron star's shock face would imply unrealistic column densities at other orbital phases. A remaining viable explanation is an eclipse of the primary by its donor star, but the data do not allow us to confirm or reject this hypothesis. Finally, we note that whatever mechanism is involved in driving the sporadic emission in SFXTs could be responsible for the prolonged dip (e.g., the 33-ks dormant period of IGR~J17391$-$3021 in Fig. \ref{fig_lc}).

\subsection{IGR~J17391$-$3021}

During the initial 33 ks of the observation of IGR~J17391$-$3021, the source is in an extremely low-activity phase that is at or near quiescence (Fig. \ref{fig_lc}). The source then enters a period of enhanced activity in which the luminosity is only a factor of 5 that of the ``quiescent'' emission. \citet{Boz10} noticed similar behavior in a recent \xmm\ observation of this source. Observations prior to and after MJD 54518.92 are referred to henceforth as ``dormant'' and ``active'' epochs, respectively. The active state features 3 quasi-periodic flares separated by $\sim$15 ks intervals. The unabsorbed 0.5--10 keV luminosity is $1.3\times10^{33}$ \ergs\ ($d$/2.7 kpc)$^{2}$ when dormant, and $7.4\times10^{33}$ \ergs\ when active.

These weak flares are just above quiescence, and their peak fluxes are at the level of the faint detections (and a few upper limits) from \swift\ monitoring \citep{Rom09}. Epochs of enhanced activity just above quiescence, but well below the bright flaring episodes typical of this class, represent 60\% of all observations, so they are the most common emission state. Given more exposure time, many \swift\ upper limits would be detections suggesting that the duty cycle is even higher. Based on the source ephemeris \citep{Dra10}, which places the \xmm\ and \suz\ observations at different phases, such low-activity states are not confined to a specific part of the orbit.

%__________________________________________________________________SPEC 16207
\begin{figure*}[!t] 
\includegraphics[width=15cm,angle=0]{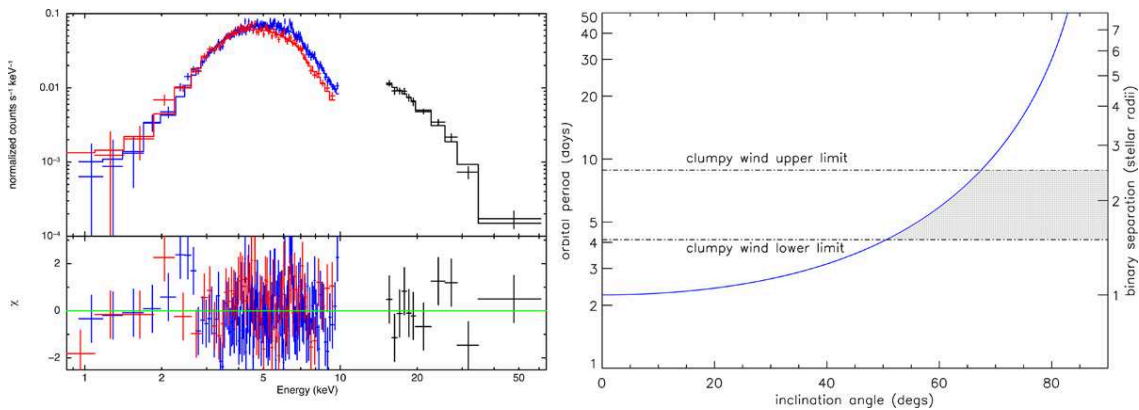}
\caption{\emph{Left}: \suz\ spectrum of IGR~J16207$-$5129 corrected for the background and fit with an absorbed power law with an exponential cutoff whose parameters are listed in Table\,1. The data represent photon counts from XIS-1 (red), XIS-0 combined with XIS-3 (blue), and HXD-PIN (black). \emph{Right}: Solutions to a model of an eclipsing SGXB \citep{Rap83} viewed under different inclination angles (degrees). The compact object is assumed to be a neutron star of 1.4\,$M_{\odot}$, and the mass of the companion star is set to 20\,$M_{\odot}$ with a radius of 20\,$R_{\odot}$. The eclipse lasts $>$30 ks (solid curve). The dot-dashed lines represent the limits of orbital radii of clumpy winds in SGXBs \citep{Neg08}. The shaded region shows the extent of the parameter space in which IGR~J16207$-$5129 might be located.}
\label{fig_spec_16207}
\end{figure*}

%__________________________________________________________________SPEC 17391
\begin{figure*}[!t] 
\includegraphics[width=15cm,angle=0]{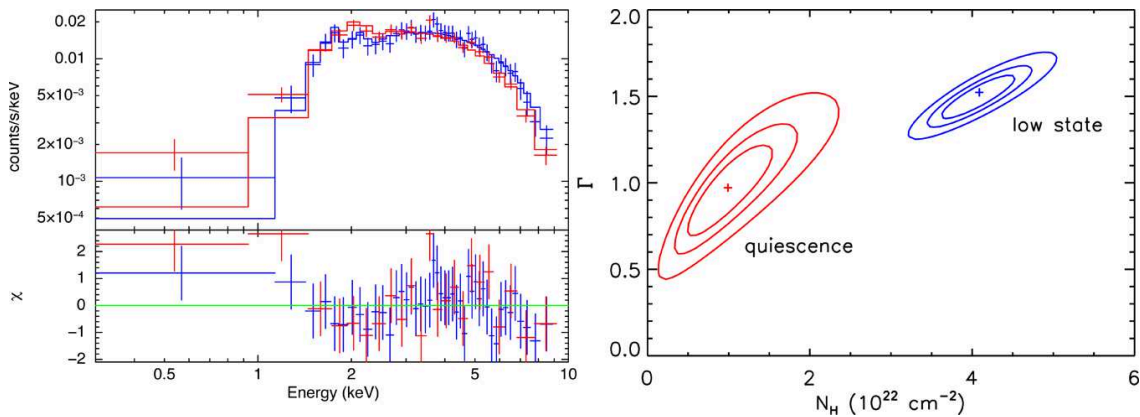}
\caption{\emph{Left}: Background-corrected spectrum of IGR~J17391$-$3021 fit with an absorbed power law (see Table\,1 for parameters). The data represent photon counts from XIS-1 (red), and XIS-0 combined with XIS-3 (blue). Each bin collects a minimum of 150 counts. \emph{Right}: Parameter space of \nh\ and $\Gamma$ derived from absorbed power laws fit to the spectra of IGR~J17391$-$3021 during dormant (red) and active (blue) states. The crosses represent the best-fitting parameters and the contours denote 68\%, 90\%, and 99\% confidence levels.}
\label{fig_spec_17391}
\end{figure*}

\section{Environment of the X-ray Emitter}

\subsection{IGR~J16207$-$5129}

The spectrum of IGR~J16207$-$5129 (Fig. \ref{fig_spec_16207}) features a large absorbing column and a possible iron line, both of which suggest the presence of matter around the X-ray source. A cutoff near 20 keV favors a neutron star as the compact object. The observed attenuation of the X-ray flux can not be explained by an increase in the column density since the measured absorption value does not change in a significant way between O2a and O2b, nor between the low and high-intensity states of O2 (Table \ref{tab_spec}). 

The eclipse interpretation remains a viable explanation and it allows us to consider the orbital configuration of the X-ray binary. An eclipse duration of at least 30 ks (0.35 d) sets a lower limit on the orbital period at 0.7 d. With this constraint, and assuming typical stellar parameters, a model for an eclipsing binary \citep{Rap83} yields a parameter space that is bounded by orbital periods between 4 and 9 d with inclination angles $>$ 50$^{\circ}$. The full orbit is within the upper limit of the clumpy-wind radius, unless the orbital period is greater than ~10 d, which is unusual for SGXBs but plausible (e.g., IGR~J19140$+$0951 has an orbital period of 14 d).

\subsection{IGR~J17391$-$3021}

Figure \ref{fig_spec_17391} presents the time-averaged spectrum of IGR~J17391$-$3021 fit with an absorbed power law. When IGR~J17391$-$3021 is in the active state, the \nh\ is at least twice as high as it is in the dormant phase. This is unlike what was seen by \citet{Boz10} who found that the \nh\ remained steady (within the statistical errors) during weak flares while the photon index varied by around 50\%. 

Differences in the spectral properties of the flares caught by \suz\ and \xmm\ could be due to unequal geometric configurations of the system between the observations. A partial-covering model fit to the \xmm\ spectrum suggests that up to 40\% of the X-ray emission is not absorbed by the clump itself, but rather by the interstellar medium. The same model fit to our active spectrum reveals an absorber that shields a much larger fraction of the X-ray source (less than 13\% of the emission is unblocked). The column density during activity is 2--4 times the interstellar value which suggests that the source is not strongly absorbed (intrinsically) except when the primary accretes a clump passing along our line of sight \citep[see also][]{Ram09}.

%__________________________________________________________________Journal
%
\begin{table*}[!t] 
%\begin{scriptsize}
\begin{tabular}{ l l l l l  l }
\hline
\hline
source			& epoch      	& \nh\				& $\Gamma$		& $L$ 				& $\chi_{\nu}^{2}/$dof	\\
				&			& $10^{22}$\,cm$^{-2}$ 	& 				& $10^{33}$\,erg\,s$^{-1}$	 & 					\\	
\hline
IGR~J16207$-$5129	& total 		& 16$\pm$1			& 0.9$\pm$0.2 		& 130				&  1.05/199		 	\\	
				& O2a  		& 19$\pm$1			& 1.3$\pm$0.1 		& 240				&  0.91/182		 	\\	
				& O2b 	 	& 19$\pm$5			& 1.5$\pm$0.4 		& 56					&  0.51/64		 	\\	
\hline
IGR~J17391$-$3021	& total 		& 3.6$\pm$0.4			& 1.4$\pm$0.1 		& 4.8					&  0.82/66		 	\\	
				& dormant 	& 1.0$\pm$0.6			& 1.0$\pm$0.3 		& 1.3					&  0.68/8		 	\\	
				& active 		& 4.1$\pm$0.5			& 1.5$\pm$0.1 		& 7.4					&  0.73/54		 	\\	
\hline
\end{tabular}
\caption{Parameters from absorbed power laws fit to the \suz\ spectra of IGR~J16207$-$5129 and IGR~J17391$-$3021 for various epochs. The luminosity in the 0.5--10\,keV band is corrected for absorption. Errors represent 90\% confidence.}
\label{tab_spec}
%\end{scriptsize}
\end{table*}

\section{Conclusions}

The putative eclipse in the persistently-emitting SGXB IGR~J16207$-$5129 awaits confirmation. As a prototypical member of the class of SFXTs, IGR~J17391$-$3021 holds valuable clues to the accretion processes of these intriguing transients. The demarcation between SGXBs and SFXTs is not as clear as originally believed, and a few systems might represent intermediate states between SFXTs and SGXBs. In this respect, the study of IGR~J16207$-$5129 and IGR~J17391$-$3021 will play an integral role in helping us understand the extent to which emissivity differences between both populations stem from their unequal wind and orbital characteristics. These results have been published as \citet{Bod10} and \citet{Bod11}.

\end{document}